\documentclass{aa}
\usepackage{graphics}
\begin{document}
\thesaurus{08(09.16.1; 09.16.2 A58 = NOVA AQL 1919; 04.19.1)}
\title{New identification of the near infrared source in the "born-again" planetary 
nebula A58 (=V605 Aql)
\thanks{Based on observations collected at the European Southern Observatory, La Silla, Chile at the 
DENIS consortium survey instrument}}
\author{S. Kimeswenger, J. Koller and S. Schmeja} 

\institute{Institut f\"ur Astrophysik der Leopold-Franzens-Universit\"at Innsbruck, 
Technikerstra{\ss}e 
25, A-6020 Innsbruck, AUSTRIA\protect\newline
http://astro.uibk.ac.at}
\offprints{S. Kimeswenger\protect\newline Stefan.Kimeswenger@uibk.ac.at}
\date{Received 25 May 2000 / Accepted 6 June 2000}
\titlerunning{New identification of the NIR source in A58}
\maketitle

\begin{abstract}

The central dust knot V605 Aql (= Nova Aql 1919) in the planetary nebula A58 is the result of a 
late-helium flash. Previous near infrared (NIR) measurements yield very bright fluxes.
This exceeds model predictions, based on other wavelengths and estimates of the visual extinction 
within the object, by a factor of 100. Using NIR imaging 
by the DENIS instrument, we found that the source was misidentified 
in previous (single channel photometer) observations. We present the NIR flux of A58 and 
identify an unrelated stellar field source, which in fact represents the NIR source attributed to A58,
according to the literature. 
The new identification is consistent with model predictions for the source. An accurate
astrometry for the core of A58 is also provided.

\keywords{(ISM) planetary nebulae: general - planetary nebulae 
(individual: A58 = V605 Aql = Nova Aql 1919) - surveys}
\end{abstract}

\section{Introduction}
In 1919 the central star of the planetary nebula A58 underwent the rare event of a 
very late helium flash ("born-again" scenario; Iben et al. \cite{Ib83}; Iben \cite{Ib84}). 
For a review of the object see Guerrero \& Manchado (\cite{Ma96}) and Clayton \& de Marco (\cite{Cl97}).
It is only one out of two such events in modern astronomy (the second event was 
Sakurai's "Novalike Object" in 1996; D\"urbeck \& Benetti \cite{Du96})
Those objects undergo strong phases of dust condensation (Kimeswenger et al. \cite{Ki97}). 
Thus, several infrared measuring campaigns focused on this object. While the mid infrared domain was 
observed with the imaging devices of the ISO satellite (Kimeswenger et al. \cite{Ki98a}), in the near 
infrared (NIR) only single channel 
aperture photometry was available (Harrison \cite{Ha96}; van der Veen et al. \cite{Ve89}). The latter 
authors give 
J = 10.25, H = 9.32, K = 9.05 and J = 10.32, H = 9.32, K = 9.10 respectively. However,
their measurements suffer from low angular resolution and pointing accuracy of the 
telescopes. Van der Veen et al. provide (Tab. 5 therein) typical errors of 15\arcsec-20\arcsec\ 
and maximum errors up to 40\arcsec\ for the positional deviation between their NIR source and 
the IRAS source. Even if they interpret this mainly as an error caused by IRAS,
it contains also a significant contribution from the NIR position. Modelling the complete spectral 
energy distribution of this object (Koller \& Kimeswenger \cite{Ko99}, \cite{Ko00a}, \cite{Ko00b}), 
the near infrared excess was typically a factor of 100. This is consistent with models of 
similar objects like V4334 Sgr (= Sakurai, Kerber et al. \cite{Ke99}). While Harrison (\cite{Ha96}) tried 
to describe this excess by a second component of extremely hot dust, Kimeswenger et al. (\cite{Ki98a}) 
tried to explain it as the hot central star obscured by several magnitudes
of extinction (Seitter 1987). The latter model did not hold, as the luminosity of such an 
object exceeds by far the complete measured bolometric luminosity. To solve this problem, 
we obtained images at 0.8$\mu$m, 1.2$\mu$m and 2.15$\mu$m using the DENIS instrument 
(Epchtein et al. \cite{Ep94}, \cite{Ep97}) to investigate the exact location of the NIR emission. 
We were able to find the real flux associated with this source and to identify the 
NIR source misinterpreted as V605 Aql by Harrison (\cite{Ha96}) and van der Veen et al. (\cite{Ve89}).

\section{Results and Discussion}
The data were obtained using the DENIS survey instrument at the ESO 1m telescope 
(Epchtein et al. \cite{Ep97}, Kimeswenger et al. \cite{Ki98b}) on May 7$^{th}$, 2000 8:15 UT. 
The images were taken simultaneously in all three bands, Gunn-I (0.82$\mu$m),
J (1.2$\mu$m) and Ks (2.15$\mu$m). Each band was observed with five images while moving 
around the source in the field of view. This was used to eliminate 
errors due to local flatfield effects and to be able to obtain the sky background using 
the iso-airmass median sky filtering. This procedure provides also an estimate of the 
intrinsic error of the measurement, being better than 0.03 mag.
The flux calibration was done individually using all standards of the night. We did not 
use the extinction correction applied for the DENIS survey (Epchtein et al \cite{Ep99}; 
Fouqu{\'e} et al. \cite{Fo00}) but derived it individually for all three bands. 
Fig. 1 shows the first images of the sequence in all bands. The optical identification 
was done using the digital sky survey (Fig. 2) and the plate presented by 
Pollacco et al. (\cite{Po92}). The near infrared flux for V605 Aql (position marked by a cross) 
can be given only as upper limits (J $<$ 15.8, Ks $<$ 13.5).
In the red visual band (Gunn-I) it is just detected near the limit 
($\approx$  18.0$\pm$0.25). 

\begin{figure}
\resizebox{\hsize}{!}{\includegraphics{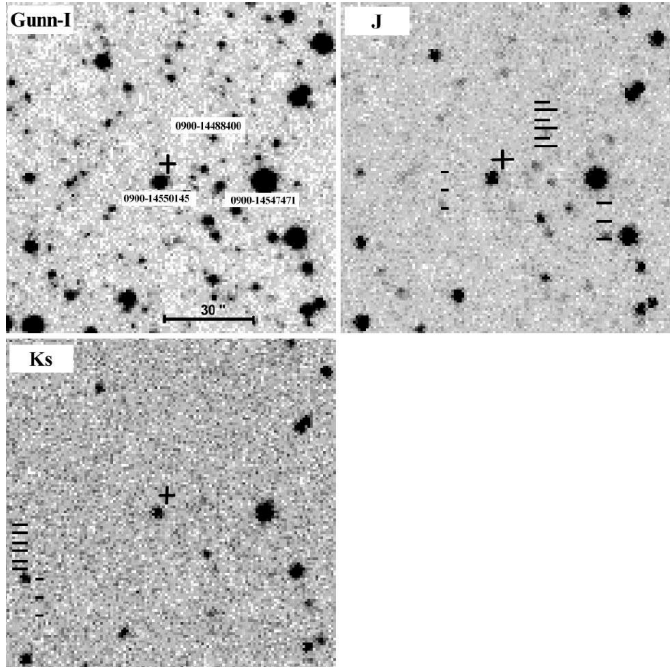}}
\caption{The DENIS NIR images of the region around A58 (V605 Aql). 
The position of the (weak) source is indicated by 
a cross in all three bands. The numbers indicate the USNO A2 
star numbers of the astrometric reference frame used. The sets of horizontal 
black lines in the NIR images are detector defects.}
\end{figure}

The astrometry was obtained using both, the red Digital Sky Survey 2 image (DSS2)
and the DENIS images. The stars USNO A2 0900-14547471, 0900-14550145 and 0900-14488400 
were used as reference and the next nearby TYCHO2 sources (about 3\arcmin\ distance) 
were used as a control frame to estimate the USNO A2 errors in that region.
There is a weak indication for a systematic shift (using the next nearby 
TYCHO stars USNO A2 0900-14571943, 0900-14569482, 0900-14571215, 0900-14529771 and 
0900-14582492) of $\Delta\alpha$ = -0.24\arcsec, $\Delta\delta$ = 0.15\arcsec.
As the rms noise is about 0.40\arcsec, we did not apply this shift to the resulting 
coordinates.

\begin{figure}
\resizebox{\hsize}{!}{\includegraphics{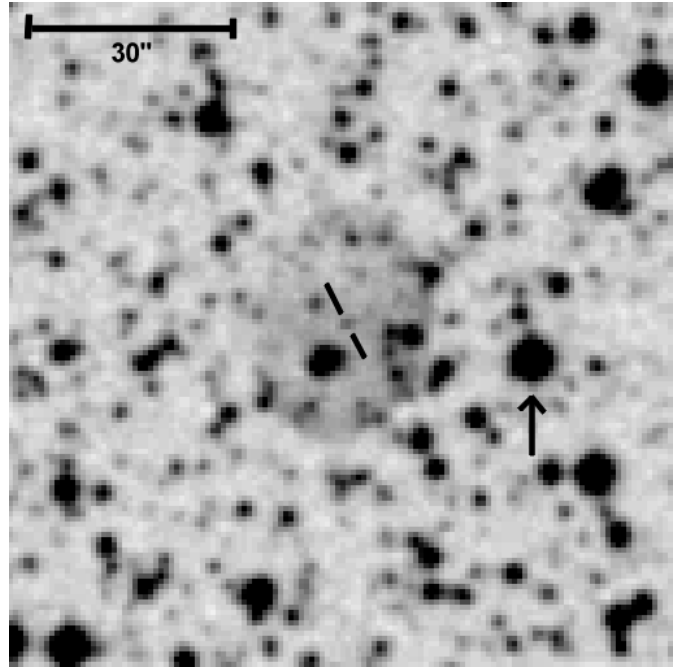}}
\caption{The red DSS2 image of the region arround A58 (V605 Aql). 
The position of the "born-again" core of A58 is indicated near the center.
The arrow marks the star USNO A2 0900-14547471, which is most likely the 
NIR source measured by Harrison (\cite{Ha96}) and van der Veen et al. 
(\cite{Ve89}).}
\end{figure}
 
Although star 0900-14550145 has the smallest distance to the core of A58, 
it is not usable for precise 
\hyphenation{cooridnates} coordinates. 
There are clearly two weak stars blended on the DSS2. Thus, the coordinates for 
A58 were obtained using only the two other nearby stars mentioned above. 
The coordinates then were measured both, on the DENIS frames and
on the DSS2, independently. The accuracy (rms) is better than 0.5{\arcsec}
relative to the USNO A2 reference frame.

\[ \begin{array}{c c l}
\alpha_{A58} & = & \mbox{\tt19}^{h}\mbox{\tt18}^{m}\mbox{\tt20}{\fs}\mbox{\tt42}~~\mbox{(J2000)} \\
\delta_{A58} & = & \mbox{\tt01}\degr\,\mbox{\tt47}\,\arcmin\,\mbox{\tt01}{\farcs}\mbox{\tt1}~~~\,\mbox{(J2000)}\\
\end{array}\]

The star marked in Fig. 2 by the arrow 34.9\arcsec\ right from the nebula  
core is USNO A2 0900-14547471 and has a 
brightness of Gunn-I = 11.73, J = 10.19, Ks = 9.08. As there is no other source 
as bright and having a similar color 
(J-K = 1.11), it is most likely the source measured by Harrison 
(\cite{Ha96}; J = 10.25, K = 9.05) 
and van der Veen et al. 
(\cite{Ve89}; J = 10.32, K = 9.10). Also the distance provided by van der Veen et al. 
supports this result. 

Using the model in Koller \& Kimeswenger (\cite{Ko00b}) we predict Gunn-I = 16.9 , 
J = 15.0, Ks = 13.0. Although this is a little bit brighter than the measured 
flux, it is well consistent with the magnitudes/limits given above.
The models show that at those NIR wavelengths already $\approx$70$-$80{\%} 
of the radiation originate from the gas component. The input parameters for the gas
are still very uncertain due to the lack of good optical spectra 
as the gas free-free emission, 
contributes already up to 80\% in those wavelengths. 

\begin{acknowledgements}
This  project  was  supported by the Austrian FWF  project P11675-AST.
We thank the DENIS consortium (PI N. Epchtein, Nice, F) for being able
to do these additional observations. The DENIS project  is
partly funded by European Commission grants. It is also supported 
in  France
by  the INSU, CNRS,
in Germany  by the State of Baden-W\"urttemberg, in Spain by the DGICYT,
in  Italy by the CNR, in Austria by
the  Fonds  zur  F\"orderung  der wissenschaftlichen  Forschung  und
Bundesministerium f\"ur Wissenschaft und Forschung,  in  Brazil  by
the  FAPESP, and  in Hungary by an OTKA   grant
and an  ESO C \& EE  grant.
\end{acknowledgements}

\end{document}